\documentclass[12pt,preprint]{aastex}

\shorttitle{$Fermi$ observations of high-energy gamma-ray emission from GRB~090217A}

\usepackage{natbib}
\begin{document}

\title{$Fermi$ observations of high-energy gamma-ray emission from GRB~090217A}
\author{
M.~Ackermann\altaffilmark{2}, 
M.~Ajello\altaffilmark{2}, 
L.~Baldini\altaffilmark{3}, 
J.~Ballet\altaffilmark{4}, 
G.~Barbiellini\altaffilmark{5,6}, 
M.~G.~Baring\altaffilmark{7}, 
D.~Bastieri\altaffilmark{8,9}, 
K.~Bechtol\altaffilmark{2}, 
R.~Bellazzini\altaffilmark{3}, 
B.~Berenji\altaffilmark{2}
% ...
}
% ...
\altaffiltext{2}{W. W. Hansen Experimental Physics Laboratory, Kavli Institute for Particle Astrophysics and Cosmology, Department of Physics and SLAC National Accelerator Laboratory, Stanford University, Stanford, CA 94305, USA}
\altaffiltext{3}{Istituto Nazionale di Fisica Nucleare, Sezione di Pisa, I-56127 Pisa, Italy}
\altaffiltext{4}{Laboratoire AIM, CEA-IRFU/CNRS/Universit\'e Paris Diderot, Service d'Astrophysique, CEA Saclay, 91191 Gif sur Yvette, France}
\altaffiltext{5}{Istituto Nazionale di Fisica Nucleare, Sezione di Trieste, I-34127 Trieste, Italy}
\altaffiltext{6}{Dipartimento di Fisica, Universit\`a di Trieste, I-34127 Trieste, Italy}
\altaffiltext{7}{Rice University, Department of Physics and Astronomy, MS-108, P. O. Box 1892, Houston, TX 77251, USA}
\altaffiltext{8}{Istituto Nazionale di Fisica Nucleare, Sezione di Padova, I-35131 Padova, Italy}
\altaffiltext{9}{Dipartimento di Fisica ``G. Galilei", Universit\`a di Padova, I-35131 Padova, Italy}
% ...

\begin{abstract}
The $Fermi$ observatory is advancing our knowledge of Gamma-Ray Bursts (GRBs) 
through pioneering observations at high energies, covering more than 7 decades in energy with 
the two on-board detectors, the Large Area Telescope (LAT) and the Gamma-ray Burst 
Monitor (GBM).
Here we report on the observation of the long GRB~090217A which triggered the GBM and has been
detected by the LAT with a significance greater than 9~$\sigma$.
We present the GBM and LAT observations and on-ground analyses, including the time-resolved spectra
and the study of the temporal profile from 8~keV up to $\sim$1~GeV. 
All spectra are well reproduced by a Band model.
We compare these observations to the first two LAT-detected, long bursts GRB~080825C and GRB~080916C.
These bursts were found to have time-dependent spectra and exhibited a
delayed onset of the high-energy emission, which are not observed in the case of GRB~090217A.
We discuss some theoretical implications for the high-energy emission of GRBs.
\end{abstract}



\section{Introduction}
Gamma-Ray Bursts (GRBs) are the most extreme events in the Universe, and have fascinated astronomers since their discovery forty years ago \citep{klebe}. 
Until 2008, only a small fraction of GRBs had been detected at energies above the MeV region (Hurley et al. 1994, Gonzalez et al. 2003, Giuliani et
al. 2008).
The $Fermi$ Gamma-ray Space Telescope was launched on June 11, 2008, and provides unprecedented energy coverage and sensitivity for the study of
high-energy emission in GRBs.
It is composed of two instruments: the Gamma-ray Burst Monitor (GBM) \cite{meegan09} and the Large Area Telescope (LAT) \cite{atwood09}.
The GBM monitors the entire unocculted sky. It covers four decades in energy through the combination of 12 NaI and 2 BGO scintillation detectors, which are
sensitive in the energy ranges 8~keV to 1~MeV and 150~keV to 40~MeV, respectively.
The LAT is a pair conversion telescope which consists of $4 \times 4$ arrays of silicon strip trackers and cesium iodide calorimeter modules covered by a
segmented anti-coincidence detector designed to efficiently reject charged particle background events. 
The energy coverage of the LAT instrument ranges from 20~MeV to more than 300~GeV, with a field of view of $\sim$2.4~sr at 1~GeV.  

After 18 months of operations, the LAT instrument has detected 15 GRBs at energies above 100~MeV, including two short bursts\footnote{See the LAT GRB table: 
http://fermi.gsfc.nasa.gov/\-ssc/\-resources/\-observations/\-grbs/\-grb\_table}.
After GRB~080825C \cite{080825c} and GRB~080916C \cite{080916c}, GRB~090217A is the third long burst which was firmly detected with the LAT above 100~MeV after
triggering the GBM.
Section~\ref{sec:observations} addresses the detection and localization of this burst, and presents the light curves from both instruments.
The joint spectral analysis is described in section~\ref{sec:jointAnalysis}.
In section~\ref{sec:discussion} the results are discussed in the context of other LAT-detected long GRBs and current theoretical models, and our conclusions are
given in section~\ref{sec:conclusion}.

\section{GBM and LAT observations}
\label{sec:observations}

\subsection{GBM observations}
\label{subsec:GBMobservations}

At 04:56:42.56 UTC on February 17$^{th}$ 2009 ($T_0$), the $Fermi$ GBM triggered and located GRB~090217A (trigger 256539404 / 090217206) \cite{GCN_8902}.
This bright, long GRB was detected by NaI detectors numbers 6 to 11 which are located on one side of the spacecraft.
Four of these detectors
exceeded the threshold and the GBM triggered on 256 ms timescale in the energy range 50 to 300 keV.
The BGO detector located on the same side, B1, detected emission up to $\sim$1~MeV.
In fact, the emission from GRB~090217A was so intense that it was detected through the spacecraft by most of the NaI detectors and indeed
the BGO detector (B0) on the opposite side.\\

The angle of GRB~090217A to the LAT boresight was 35$^\circ$, placing this bright, hard event firmly in the field of view.
The GBM on-ground localization reported by von Kienlin (2009) was unusually far ($\sim$9$^\circ$) from the early LAT localization \cite{GCN_8903}.
However, this was improved upon using a response table for spectrally hard, bright bursts such as GRB~090217A which also accounts for saturation events.
The final GBM position is (RA, Dec) = ($205.5^\circ$, $-6.0^\circ$) with a statistical error of $\sim$1$^\circ$ and a standard systematic uncertainty of
2$^\circ$--3$^\circ$ \cite{briggs09}, and is consistent with the LAT position.\\

The light curves from the GBM and LAT are shown in Fig.~\ref{Fig:LC}, sorted from top to bottom in order of increasing energy.
The first two panels display the background-subtracted light curves for the NaI and the BGO detectors.
The first panel shows the sum of the counts of the three NaI detectors (N6, N8 and N9) with the strongest signal in the 8--260~keV band.
The second panel shows the corresponding plot for the sum of both BGO detector counts, between 260~keV and 5~MeV.
GRB~090217A features a structured light curve in the GBM energy band, with one major peak with substructures and many
overlapping pulses.
The burst duration estimates for the 8~keV to 1~MeV range are T$_\mathrm{90}$$=$32.8~s, and T$_\mathrm{50}$$=$11.3~s.
The LAT light curves in the last three panels are described in section~\ref{subsubsec:LATroiE}.

\subsection{LAT observations}

\subsubsection{Detection and localization}
\label{subsubsec:LATdetection}
GRB~090217A did not trigger the LAT onboard detection algorithm and was found through a blind search in the LAT data by the on ground Automated Science Processing
(ASP) pipeline \cite{band09}. 
Subsequently and independent of the ASP analysis, the detection was confirmed by searches of the GBM location.
The first step of this study is selecting the LAT events belonging to the so-called `TRANSIENT' class \cite{atwood09} of the `P6\_V3'
analysis\footnote{http://www-glast.slac.stanford.edu/software/IS/glast\_lat\_performance.htm}.
This selection provides a large effective area with a reasonable background rate adapted for burst detection and localization.\\


The unbinned likelihood method for localization makes use of the LAT Point Spread Function (PSF) on an event by event basis.
The instrument response functions (IRFs) have not been validated below 100~MeV, thus we restrict this analysis to be above this spectroscopic threshold.
Following the methodology described in detail in \cite{080825c}, we computed the LAT position of GRB~090217A using the map of the Test Statistics (TS),
considering all TRANSIENT events recorded above 100~MeV between $T_0$ and $T_0$+37.5~s in a region of 15$^\circ$ centered on the final GBM position.
GRB~090217A occurred at a relatively high Galactic latitude ($b$$\sim$53$^\circ$), thus the Galactic emission contributes only a few percent of the total
background. Therefore, only an isotropic component, largely residual charged-particle events,
was included in the background model.

To compute the detection significance and the localization error, the TS values are interpreted in terms of the $\chi^2$ distribution with two degrees of
freedom.
The best fit position is found to be (RA, Dec) = ($204.73^\circ$, $-8.43^\circ$), which is $0.17^\circ$ away from the early localization \cite{GCN_8903}, with
a TS$_\mathrm{max}$=89 corresponding to a 9.2~$\sigma$ detection.
The TS contours around this position yielded the 68\%, 90\% and 99\% statistical error radii, respectively of $0.37^\circ$, $0.54^\circ$ and $0.80^\circ$.
As explained in \cite{080825c}, the relatively small inclination angle of GRB~090217A in the LAT field of view implies a negligible systematic error
($<$0.1$^\circ$).\\

We also searched for a temporally extended emission above 100~MeV, as observed in other LAT long bright bursts, e.g. GRB~080916C \cite{080916c}
and GRB 090902B \cite{090902b}.
This unbinned likelihood analysis is based on the LAT events belonging to the so-called `DIFFUSE' class, which is
suited to searches on longer time scales \cite{080825c}.
GRB~090217A remained within 60$^\circ$ from the LAT boresight until $T_0$+500~s, but no additional signal was found by this time.
Finally, despite Swift Target of Opportunity observations of the early LAT localization, no X-ray afterglow was found \cite{GCN_8907}
and hence there is no redshift available for this burst.

\subsubsection{Count light curves from energy-dependent spatial event selections}
\label{subsubsec:LATroiE}

Unlike the unbinned likelihood analysis, the joint GBM-LAT spectral study and the study of the burst temporal profile in the LAT do not make use of the
LAT PSF on an event by event basis.
In these analyses, an increased signal-to-background ratio is obtained for the LAT by selecting the events in a region of interest (ROI) centered on the final LAT
position.
As described in \cite{080825c}, its size ($r_\mathrm{ROI}$) is energy dependent, and is obtained as the $95\%$ containment radius of the LAT PSF added in
quadrature to the LAT localization error.

\begin{figure}[t!]
\includegraphics[width=\linewidth]{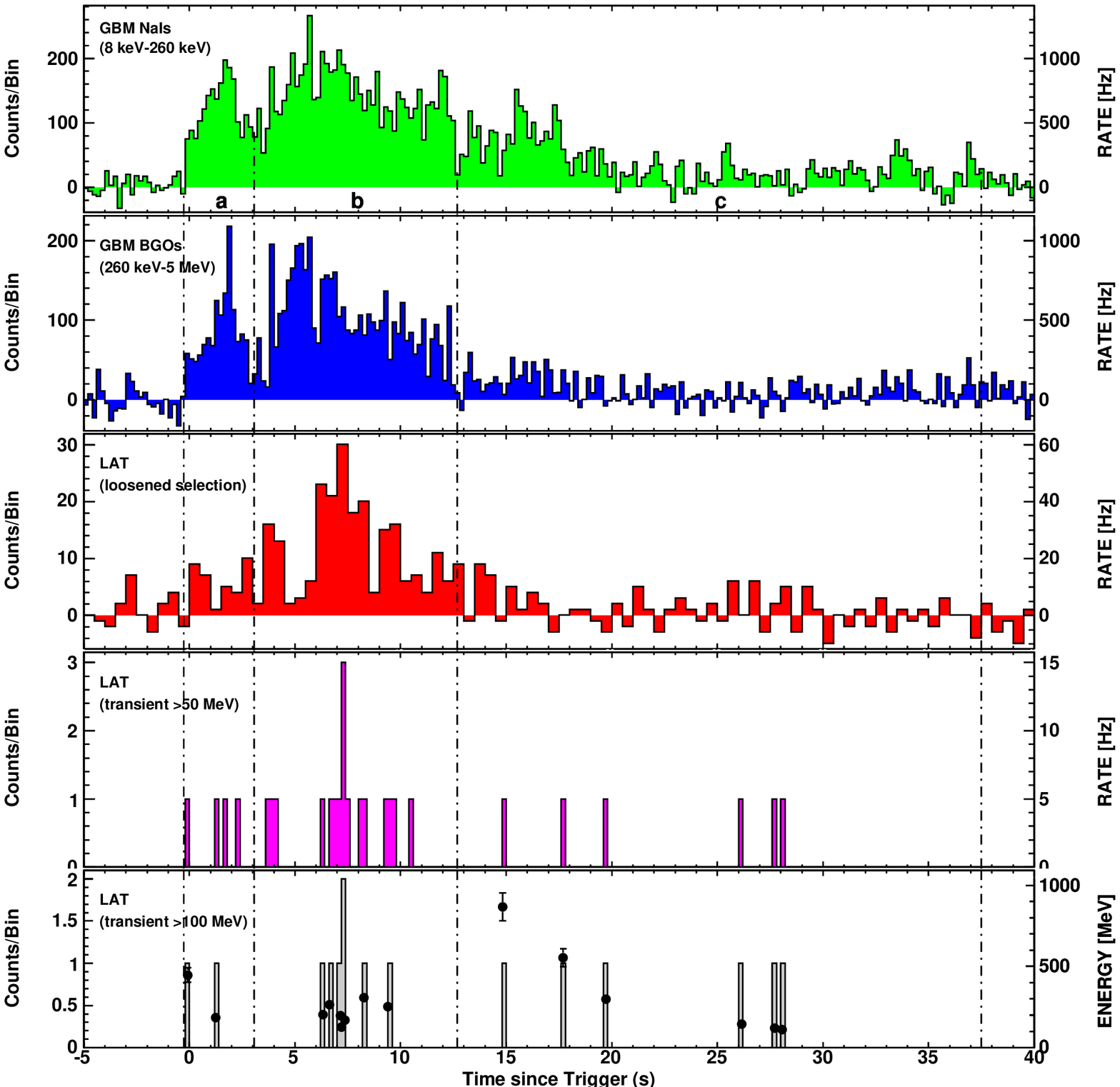}
\caption{Light curves for GRB~090217A observed by the GBM and the LAT, from lowest to highest energies. The top three panels are background subtracted.
The first panel shows the sum of the counts of the three NaI detectors (N6, N8 and N9) with the highest signal in the 8-260~keV band.
The second panel shows the sum of both BGO detector counts, between 260~keV and 5~MeV.
The LAT light curve shown in the third panel has been generated using events which passed the onboard gamma filter, with a direction compatible
with the final LAT position (see text). The binning is 0.5~s, while it is 0.2~s in the other panels.
The last two panels have been generated from the LAT events which passed the TRANSIENT event selection above 50~MeV
and 100~MeV, respectively. The events were selected within a ROI centered on
the final LAT position, with a size decreasing with energy following the PSF energy dependence (see text).
Black dots, along with their error bars (uncertainty in the LAT energy measurement) represent the 1-$\sigma$ energy range (right y-axis) for each LAT
event in the last panel. The vertical dash-dotted lines indicate the time bins used in the time-resolved spectral analysis.}
\label{Fig:LC}
\end{figure}

We split the LAT TRANSIENT events into `FRONT' and `BACK' data sets, respectively including the events which converted in the upper and lower part of the
tracker, thus with different PSF widths \citep{atwood09}.
We used a maximum size $r_\mathrm{ROI}^\mathrm{max}$=10$^\circ$--12$^\circ$ for these two conversion types, which makes the size of the ROI
saturate between 100~MeV and 200~MeV in both cases.
All events resulting from this selection process are shown in the last two panels of Fig.~\ref{Fig:LC}.
The events above 100~MeV will be used for the joint GBM-LAT spectral analysis in section~\ref{sec:jointAnalysis}.
They are displayed in the last panel along with their energies as a function of time.
A 447~MeV event is detected at $\sim$$T_0$ while the LAT event with the highest energy (866~MeV) arrives at $T_0$+14.8~s.
Additional events are recorded up to $\sim$30~s after the trigger time.\\

\begin{figure}[t!]
\hbox{
\includegraphics[width=\linewidth]{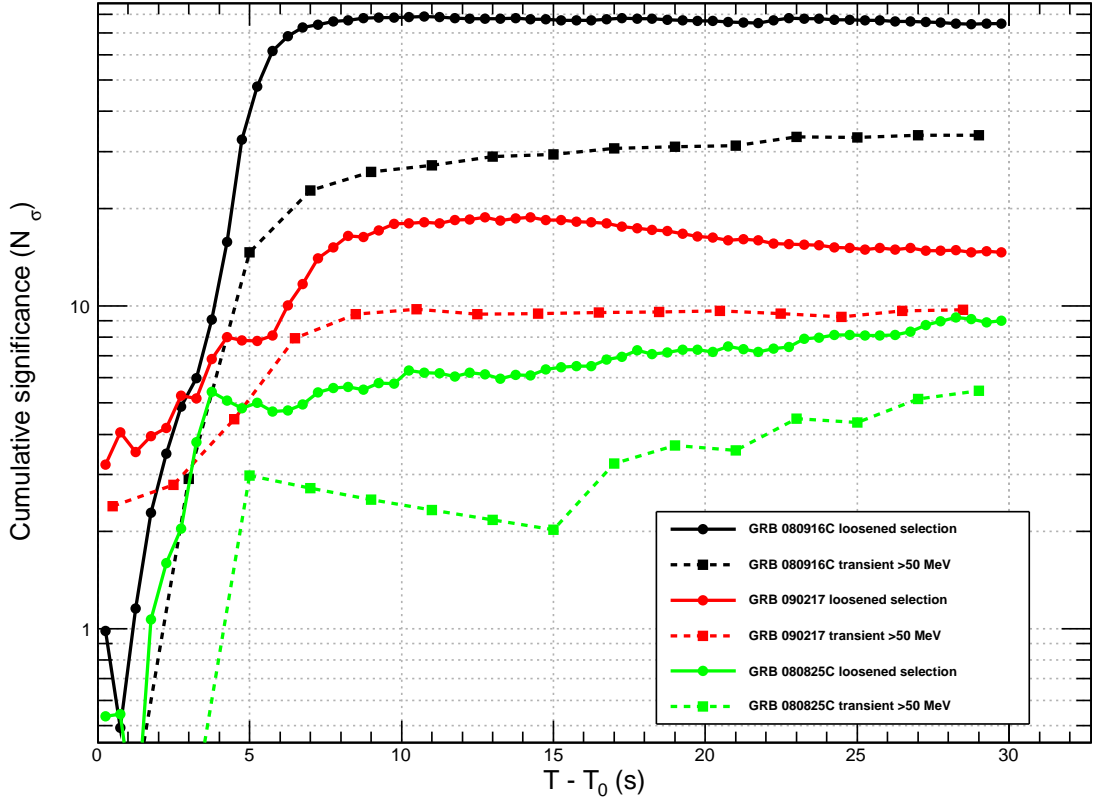}
}
\caption{Time history of the cumulative significance of the LAT emission for GRB~090217A, using the loosened selection and the event
  selection above 50~MeV shown in the third and fourth panels in Fig.~\ref{Fig:LC}.
Similar curves for GRB~080825C and GRB~080916C are superimposed.}
\label{Fig:Probs}
\end{figure}

The study of the temporal profile of the LAT emission which is discussed in section~\ref{sec:discussion} requires good photon statistics.
In the case of GRB~090217A, this can be achieved by including events recorded at lower energies, i.e. the 27 TRANSIENT events above 50~MeV which are
displayed in the fourth panel of Fig.~\ref{Fig:LC}.
As in \citep{080825c}, the expected number of background events is computed by a procedure developed by the LAT collaboration to quantitatively estimate levels
of residual charged-particle backgrounds. This estimator
was found to provide accurate values of the TRANSIENT background level above 50~MeV within a 10\%--15\% systematic uncertainty.
The computation of
the probability that the observed number of counts is due to a background
fluctuation is performed in a frequentist way, with a semi Bayesian treatment of the systematic uncertainty.
The time history of the corresponding cumulative significance for the gamma-ray signal is shown in Fig.~\ref{Fig:Probs}.

As in \cite{081024b}, we also used a relaxed event selection, considering all LAT events that passed the onboard gamma filter.
Most of these events
have at least a well-reconstructed track in the tracker, thus providing a
rough direction measurement.
The corresponding PSF was found to be much worse than for the TRANSIENT class, with a 68\% containment radius of $\sim$20$^\circ$, $\sim$13$^\circ$ and
$\sim$7$^\circ$ at 20~MeV, 50~MeV and 100~MeV, respectively.
However, applying an additional spatial selection based on the 68\% containment angles for this PSF reduced the background from $\sim$300~Hz to $\sim$16~Hz.
The background-subtracted light curve obtained with this loosened event selection is shown in the third panel of Fig.~\ref{Fig:LC}, and the time history of its
cumulative significance is reported in Fig.~\ref{Fig:Probs}.


\section{GBM and LAT joint spectral analysis}
\label{sec:jointAnalysis}

Simultaneous spectral fits of the GBM and LAT data were performed for each of the three time bins (a to c) shown in Fig.~\ref{Fig:LC}.
The selected boundaries reflect the time characteristics of GBM features and LAT photons.
The first time bin (a) starts 256~ms before the GBM trigger time in order to include the first LAT (447~MeV) event, and lasts up to $T_0$+3.072~s.
The central time bin (b) starts at $T_0$+3.072~s up to $T_0$+12.672~s, i.e. until the end of the main LAT emission.
The last bin (c) starts at $T_0$+12.672~s and ends at $T_0$+37.504~s, covering the tail of the LAT and GBM emission.
The 15 total events selected in the LAT are displayed in the last panel of Fig.~\ref{Fig:LC}.
As in section~\ref{subsubsec:LATdetection}, we used a spectroscopic threshold of 100~MeV to avoid any spurious result due to systematic uncertainties in the LAT
IRFs.
The spectral analysis was performed with the software package RMFIT (version 3.1), using binned GBM TTE data and selected LAT FRONT and BACK events  \citep[see details in][]{080825c}.
Instead of a $\chi^2$, we used the Castor C-statistic \cite{dorman} to simultaneously fit the combined data sets due to the small number of events at the
highest energies.
The Castor statistic is similar to the Cash statistic \cite{cash79}, except for an offset which is constant for a particular data set.

\begin{figure}[t!]
\includegraphics[width=\linewidth]{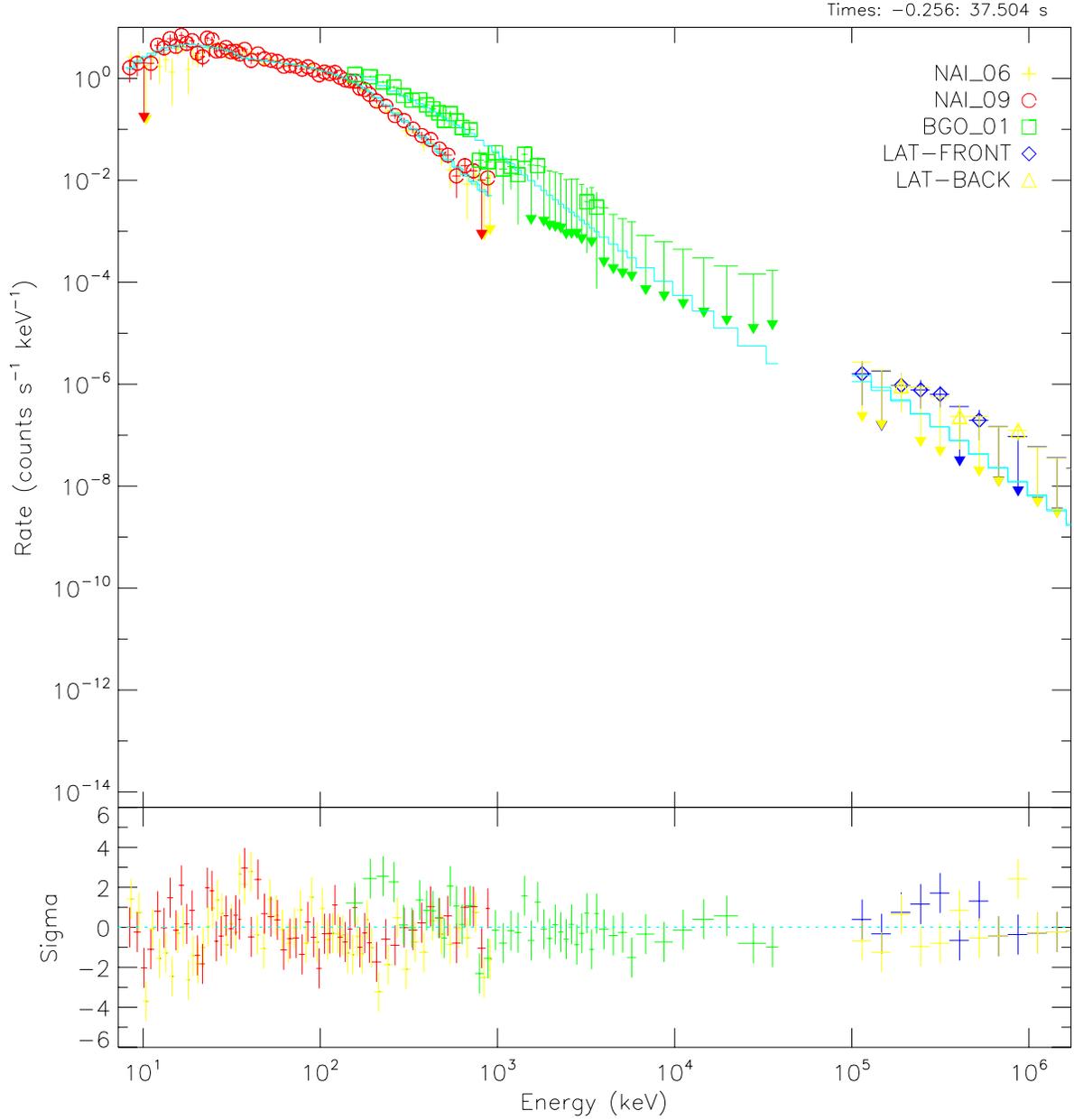}
\caption{Time-averaged ($T_0 - 0.26$~s to $T_0 + 37.50$~s) count spectrum of GRB~090217A of the GBM (NaI and BGO) and LAT data. 
The spectrum is well fit by a Band function spanning five decades in energy. The LAT data has been separated into FRONT and BACK data sets (see
text).}
\label{Fig:CountSpectrum}
\end{figure}

\begin{figure}[t!]
\includegraphics[width=\linewidth]{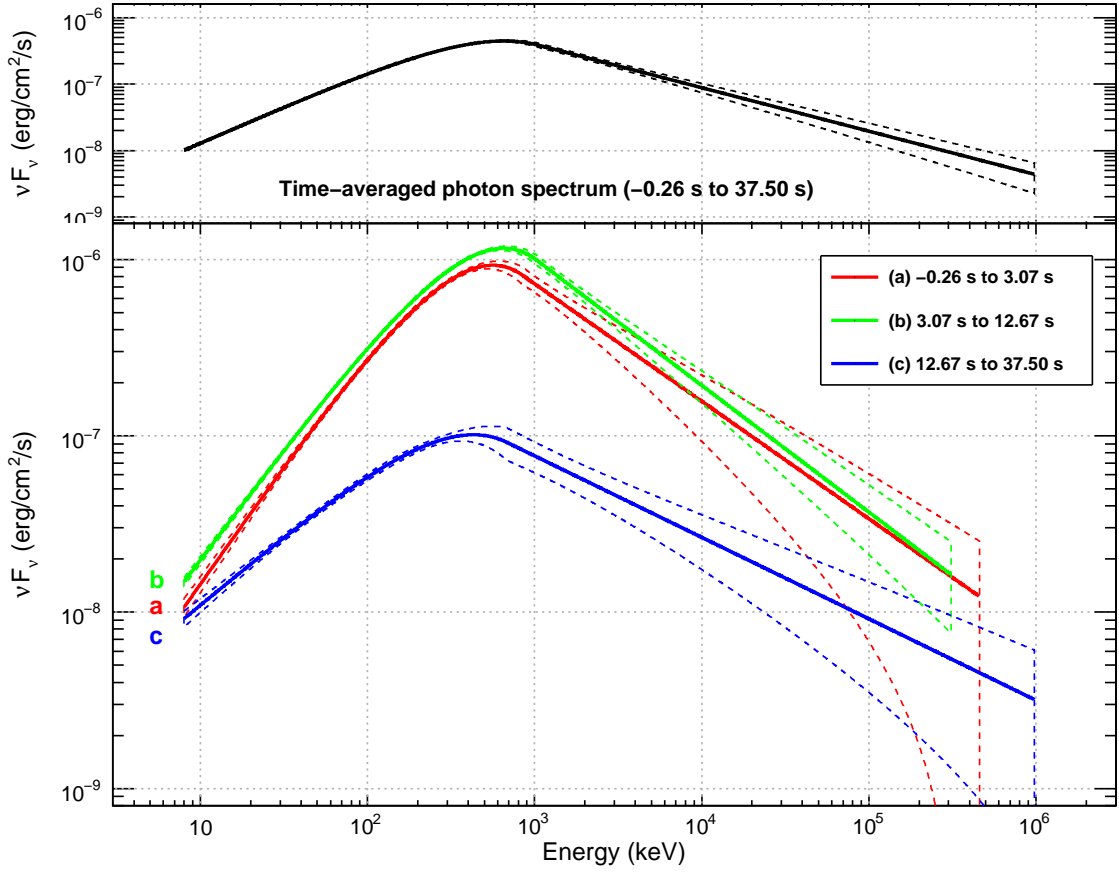}
\caption{The spectral energy distributions for the Band models found in time bins (a) to (c) are shown in thick solid lines which reach up to the largest
  detected photon energy in each time bin, while the corresponding (same color) thin dashed lines represent the 68\% confidence level contours for each fit.}
\label{Fig:NuFnu}
\end{figure}

All time-resolved count spectra are well fit by a Band function, which consists of two smoothly-connected power laws \cite{band93}.
Fig.~\ref{Fig:CountSpectrum} displays the time-averaged count spectrum,
which spans five decades in energy and is also well reproduced by a Band model.
The source photon spectral energy distributions are shown in Fig.~\ref{Fig:NuFnu} with their 68\% confidence level contours.
The best fit model parameters are reported in Table~\ref{Table:results} with their statistical errors
along with the energy fluxes in the 20~keV -- 2~MeV and 100~MeV -- 10~GeV energy bands.
The systematic errors on these parameters are essentially dominated by the uncertainty on the effective areas of the instruments.
In \cite{080825c}, we found a $\pm$15\% systematic uncertainty on the amplitude $A$, $\pm$0.03 for the spectral slopes $\alpha$ and $\beta$, and
$\pm$8~keV for the peak energy $E_{\rm{peak}}$.
For GRB~090217A, these uncertainties are negligible for $\beta$ and $E_{\rm{peak}}$, while they are comparable to the statistical errors for $\alpha$ and
dominant for $A$.

Whereas the time evolution of the amplitude and $E_{\rm{peak}}$ shows the same trend as the overall intensity of the burst, as observed from Fig.~\ref{Fig:LC},
the low-energy slope ($\alpha$) becomes gradually softer and the high-energy slope ($\beta$) remains constant within the quoted errors.
In particular, the apparent hardening of $\beta$ between time bins (b) and (c) is not significant ($\sim$1.5~$\sigma$).
The hardness ratio between the low- and high-energy fluxes is also found to be constant, with a mean value of $0.026\pm0.011$ for the time-averaged spectrum.

\section{Comparison to the LAT detected GRB~080825C and GRB~080916C}
\label{sec:discussion}
GRB~090217A does not exhibit any noticeable spectral feature, other than the common Band spectral shape,
similar to the first two long bursts detected by the LAT, GRB~080825C and GRB~080916C.
In all three bursts, no significant excess in the form of an additional spectral component, or a deficit in the form of a spectral cut off, was found in the
LAT energy range with respect to the extrapolation of the Band spectrum from lower energies.
In combination with correlated temporal behaviour of the low and high energies, this suggests that a single emission mechanism accounts for the
radiation across all energy bands.
In \cite{081024b}, we estimated the fluence ratio of the 100~MeV -- 10~GeV energy band to the 20~keV -- 2~MeV energy band to be $\sim$7\% for GRB~080825C and
$\sim$30\% for GRB~080916C. The ratio of $\sim$3\% found for GRB~090217A places this burst in a similar range as GRB~080825C.

Beyond these obvious similarities, GRB~090217A does not share several of the high-energy properties observed in the other two bursts.
First of all, the onset of the high-energy emission of GRB~090217A is not much delayed with respect to the $\sim$100~keV radiation, and a continuous increase of the
high-energy flux is observed at early times, as shown in Figs.~\ref{Fig:LC} and~\ref{Fig:Probs}.
Fig.~\ref{Fig:Probs} also shows the time history of the cumulative significance for GRB~080916C and GRB~080825C.
In spite of its brightness, GRB~080916C is not detected within the first $\sim$2 seconds, and then exhibits a very sharp rise between 3~s and 5~s.
Given the poor photon statistics from the TRANSIENT event selection, we could not confirm the presence of a lag in the case of GRB~080825C (with a
3.4\% chance probability), and simply noted that the first events recorded by the LAT from this burst were coincident with the second GBM peak \cite{080825c}.
The light curve obtained with a loosened event selection confirms the absence of any excess at early times with a higher significance.
GRB~080825C is significantly detected ($>$5~$\sigma$) only after $\sim$4~s, with the bulk of its emission slowly accumulating up to 9~$\sigma$ within the next
$\sim$25 seconds.
GRB~090217A has a smoother evolution and the delayed onset of its emission in the LAT energy band is much less marked.
It is marginally seen from the first instants and significantly detected after only $\sim$3~s.
It reaches its maximum between 10~s and 15~s and does not last longer than the low-energy emission.
This is the opposite to what was observed for the other two bursts: GRB~080916C showed evidence for a long lasting LAT emission up to 1.4~ks,
while GRB~080825C emission lasted somewhat longer in the LAT (up to $T_0$+35~s) than in the GBM (with a $T_{90}$=27~s), with the highest energy photon arriving
when the GBM emission was very weak.
In \cite{080825c}, we discussed the latter result as a possible detection of a separate and harder component showing up at late times.
Finally, no strong spectral evolution was observed in GRB~090217A especially at the highest energies, unlike GRB~080916C which underwent a strong
soft-hard-soft evolution.

\section{Discussion and Conclusions}
\label{sec:conclusion}

While other long LAT-detected bursts such as GRB~080825C and GRB~080916C exhibit a high-energy spectral variability associated to a delayed onset of the
LAT emission along with a temporally extended emission, GRB~090217A is a firmly LAT-detected burst with featureless high-energy properties.
The similarity of the temporal history of its gamma-ray emission over five decades in energy and 
the agreement, within the observational errors, between measured spectra with the Band model suggest that 
a single mechanism is responsible for the observed broad-band emission.

As in the case of GRB~080916C, a simple leptonic mechanism appears 
to be the most straightforward choice to reproduce the observed emission, 
e.g. synchrotron emission or jitter radiation \cite{med00}. The low-energy spectral 
slope of all three time bins is compatible with a synchrotron mechanism \cite{preece98},
and the mild soft to hard to soft variation of $E_{\rm{peak}}$ could be due 
to episodes of different shell collisions leading to shocks with different parameters.
More complicated scenarios are possible but not required by the present observations.
Observations of more, and brighter, GRBs with both GBM and LAT in the near future 
will certainly help to assess what fraction of high-energy emitting bursts share
similar properties, and to clarify the dominant emission mechanisms as well as 
the particle acceleration and cooling processes occurring in GRB jets.

\acknowledgments

The $Fermi$ LAT Collaboration acknowledges support from a number of agencies and institutes for both development and the operation of the LAT as well as
scientific data analysis. These include NASA and DOE in the United States, CEA/Irfu and IN2P3/CNRS in France, ASI and INFN in Italy, MEXT, KEK, and JAXA in
Japan, and the K.~A.~Wallenberg Foundation, the Swedish Research Council and the National Space Board in Sweden. Additional support from INAF in Italy and CNES
in France for science analysis during the operations phase is also gratefully acknowledged.
The $Fermi$ GBM Collaboration acknowledges support for GBM development, operations and data analysis from NASA in the US and BMWi/DLR in Germany.

\clearpage

\begin{table}[t!]
\begin{center}
\renewcommand{\arraystretch}{1.2}
\small
\begin{tabular}{c c c c c c c c c}
\hline
\multicolumn{2}{c}{Time range (s)} & $A$ & $E_{\rm{peak}}$ & $\alpha$ & $\beta$ & \multicolumn{2}{c}{Energy flux}\\
&&&&&&GBM&LAT\\
\hline\hline
(a)& -0.26--3.07 & $21.4^{\,+1.2}_{\,-1.1}$ & $562.5^{\, +57.4}_{\,-48.2}$ & $-0.63^{\,+0.06}_{\,-0.06}$ & $-2.67^{\,+0.13}_{\,-0.21}$& 216.1$\pm$8.9 &4.9 $\pm$ 5.0 \\ 
(b)&   3.07--12.67 & $23.7^{\,+0.6}_{\,-0.6}$ & $652.7^{\, +35.5}_{\,-32.8}$ & $-0.73^{\,+0.03}_{\,-0.03}$ & $-2.72^{\,+0.08}_{\,-0.11}$&267.6$\pm$5.8 & 5.0 $\pm$ 2.7 \\ 
(c)&  12.67--37.50 & $ 4.4^{\,+0.5}_{\,-0.4}$ & $430.5^{\,+125.0}_{\,-92.7}$ & $-1.20^{\,+0.08}_{\,-0.06}$ & $-2.46^{\,+0.10}_{\,-0.16}$& 30.2$\pm$2.3  &1.7 $\pm$ 1.4 \\
\hline
All & -0.26--37.50 & $ 10.5^{\,+0.3}_{\,-0.3}$ & $656.4^{\,+43.7}_{\,-39.4}$ & $-0.89^{\,+0.03}_{\,-0.03}$ & $-2.65^{\,+0.07}_{\,-0.08}$& 108.8$\pm$2.6 &2.8 $\pm$ 1.2 \\ 
\hline
\end{tabular}
\caption{Time-resolved and time-averaged spec\-tral ana\-ly\-sis re\-sults for GRB~090217A. Band function best fit parameters are
provided for all spectra.
The amplitude $A$ of the Band function is given in units of 10$^{-3}$ $\gamma$ cm$^{-2}$ s$^{-1}$ keV$^{-1}$, its peak energy $E_{\rm{peak}}$ in keV.
Energy fluxes are given in $10^{-8}$ erg cm$^{-2}$ s$^{-1}$, in the energy ranges from 20~keV to 2~MeV for the GBM and from 100~MeV to 10~GeV for the LAT.}
\label{Table:results}
\end{center}
\end{table}

\clearpage


%
\end{document}